%% file: arXiv_manuscript.tex
\documentclass[10pt]{article}
\usepackage[a4paper,top=2.0cm,bottom=1.5cm,left=2.0cm,right=1.5cm,headheight=0.5cm,headsep=0.5cm]{geometry}
\usepackage{fancyhdr}
\usepackage[square,colon]{natbib}
\usepackage{setspace}
\usepackage[dvips]{graphicx}
 \usepackage{subfigure}
 \usepackage{rotating}
 \usepackage{pstricks}
 \usepackage{ctable}
 \usepackage[pagewise]{lineno}
 \usepackage[latin1]{inputenc}
 \usepackage{multirow}
 \usepackage{natbib}
 \usepackage{amsmath}

\newcommand{\itcite}[1]{\textit{\cite{}}}

\begin{document}
\thispagestyle{empty}

\Large
\vspace*{1.8cm}
\begin{center}
  {\bf Downscaling near-surface atmospheric fields with multi-objective Genetic Programming}
  \\[0.6cm] 
\end{center}
\large
\vspace*{4.5cm}
\begin{center}
  {\sc T. Zerenner,
    V. Venema,
    P. Friederichs,
    C. Simmer}
\end{center}
\normalsize
\begin{center}
   {\it Meteorological Institute, University of Bonn\\ Auf dem Huegel 20, 53121 Bonn and Meckenheimer Allee 176, 53115 Bonn}
  
  {\it April, 2014} \\
\end{center}
\vspace*{12.0cm}
\normalsize

\normalsize

 \section*{Abstract}
{\it
The coupling of models for the different components of the Soil-Vegetation-Atmosphere-System is required to investigate component interactions and feedback processes.
However, the component models for atmosphere, land-surface and subsurface are usually operated at different resolutions in space and time owing to the dominant processes.
The computationally often more expensive atmospheric models, for instance, are typically employed at a coarser resolution than land-surface and subsurface models. 
Thus up- and downscaling procedures are required at the interface between the atmospheric model and the land-surface/subsurface models.
We apply multi-objective Genetic Programming (GP) to a training data set of high-resolution atmospheric model runs to learn equations or short programs that reconstruct the fine-scale fields (e.g., 400 m resolution) of the near-surface atmospheric state variables from the coarse atmospheric model output (e.g., 2.8 km resolution).
Like artificial neural networks, GP can flexibly incorporate multivariate and nonlinear relations, but offers the advantage that the solutions are human readable and thus can be checked for physical consistency.
Using the Strength Pareto Approach for multi-objective fitness assignment allows us to consider multiple characteristics of the fine-scale fields during the learning procedure.
}

\section{Introduction}
\noindent 
The mass and energy fluxes at the interception between soil and atmosphere significantly impact processes in the atmosphere, at the land-surface and in the subsurface.
The Transregional Collaborative Research Centre 32 on 'Patterns in Soil-Vegetation-Atmosphere-Systems' aims at a better understanding of the interplay of processes, patterns, and structures between the different components at different scales in space and time \citep{vereecken2010patterns}.
To account for interactions and feedbacks between the different components of the soil-vegetation-atmosphere system, coupled modeling systems are increasingly used (e.g., TerrSysMP by \cite{shrestha2014scale}).
When coupling different component models a scale gap occurs:
modeling land-surface and subsurface typically requires higher resolutions, while atmospheric models are often computationally too expensive to be employed at the same high resolution.
\\
The lower atmospheric boundary layer (ABL) is usually very heterogeneous due to the production of turbulent kinetic energy by shear and buoyancy effects, even more so over heterogeneous terrain.
The representation of the spatio-temporal variability in the lower ABL is very important, because the fluxes of matter and energy are driven primarily by the local differences of state variables at the surface and the lowermost atmosphere.
Many processes at the earth's surface are nonlinear, e.g., processes such as runoff production or snow melt, which are threshold dependent, or the turbulent exchange coefficients, which are nonlinear functions of the near-surface atmospheric stability \citep{schomburg2010downscaling}.
Modeling exchange processes using averaged parameters or state variables can introduce biases \citep{schlunzen2003relevance}.
To avoid such biases a downscaling algorithm is needed that reproduces the atmospheric variability near the surface and in order to provide reasonable driving data for land-surface and hydrological models.
\\
Downscaling methods have initially been developed for general circulation models (GCMs).
Various review articles summarize the large number of downscaling methods existing nowadays \citep{wilby1997downscaling, xu1999gcms,  von2000review, fowler2007linking, maraun2010precipitation}.
\\
The general idea behind downscaling is to model the variables on a smaller based on the large scale model output.
In dynamical downscaling a limited area model at high-resolution is applied.
Dynamical downscaling is computationally expensive, but it provides physically consistent spatial fields on the smaller scale.
Statistical downscaling is computationally less demanding:
regression models, computationally probably least expensive, employ a so-called transfer function between the large-scale predictors and the local-scale predictant.
Examples of regression approaches for downscaling incorporate for instance multiple linear regression, principal component analysis (e.g., \cite{wagner1990combination} or \cite{simon2013pattern}), canonical correlation analysis (e.g., \cite{friederichs2006seasonal}), artificial neural networks (e.g., \cite{schoof2001downscaling} or \cite{carreau2011stochastic}) or kriging (e.g., \cite{biau1999estimation}).
For precipitation downscaling also fractals are used (e.g., \cite{deidda2000rainfall}).
Weather generators (WGs) can be seen as an expansion of the regression models, as WGs not only try to estimate the mean but also the variance. 
Finally, weather typing or classification schemes relate synoptically or statistically  defined weather classes to the local climate.
The local variables are modeled conditioned on the weather classes.
\\
\cite{fiddes2014toposcale} recently introduced TopoSCALE to downscale GCM data to scales of $<$ 100 m. 
TopoSCALE is a physically based downscaling scheme designed for regions with complex terrain, where vertical gradients of the atmospheric variables often dominate over horizontal gradients.
A recent study by \cite{malone2012general} presents a downscaling methodology where the uncertainty of the target variable is taken into account when fitting a nonlinear regression model.
\\
In general, downscaling of meso-scale model output is much less common than the downscaling of GCM output,
but most concepts on the regional climate scale can be adapted to the smaller scales.
As we aim at a downscaling algorithm applicable within a coupled modeling system, methods with low computational costs are preferable.
\cite{schomburg2010downscaling, schomburg2012disaggregation} have presented a three-step downscaling scheme combining a bi-quadratic spline-interpolation, deterministic rules based on linear regression and autocorrelated noise.
The scheme works well for certain variables and weather conditions, but a linear regression appears to be insufficient under more complex conditions.
In this study we apply an advanced regression approach taking into account multivariate and nonlinear relations.
Artificial neural networks (ANNs) are widely used for downscaling, but we take an alternative approach, Genetic Programming (GP).
Like ANNs, GP can handle multivariate and nonlinear relations, but with the advantage that the solutions take the form of an equation or program code, which is human readable and thus can be checked for physical consistency.
\\
GP is a machine learning method from the area of evolutionary computation  \citep{banzhaf1997genetic, koza1992genetic}.
As the name implies such methods are inspired by the concept of natural evolution.
Given a training data set and a fitness measure that quantifies the quality of a candidate solution (i.e., a downscaling rule) GP automatically generates solutions.
GP typically starts by randomly generating an initial population of candidate solutions.
Each following generation is created by mutating and recombining the candidate solutions from the previous generation with the best solutions most likely contributing to the new generation.
\\
In atmospheric and related sciences GP is applied rather rarely.
\cite{liong2002genetic}, for instance, applied GP for rainfall-runoff modeling.
\cite{parasuraman2007modelling} and \cite{izadifar2010prediction} employed GP for modeling evapotranspiration or latent heat from meteorological variables.
\cite{ghorbani2010sea} used GP for sea water level forecasting.
In these four studies GP performed reasonably well.
Three of these studies compared the performance of GP and ANNs.
Both methods preformed about equally well, which underlines that GP is an eligible alternative to ANNs, which are more popular in geosciences.
\\
Only very few studies apply methods from evolutionary computation to atmospheric downscaling.
To our current knowledge all of them deal with downscaling of GCM output to a station or catchment mean.
\cite{coulibaly2004downscaling} utilized GP for simulating daily extreme temperature at a weather station located in the Chute-du-Diable basin in Northeastern Canada.
GP yielded good results, outperforming the widely used Statistical Down-Scaling Model (SDSM) by \cite{wilby2002sdsm}.
\citet{hashmi2011statistical} employed Gene Expression Programming (GEP), which is a variant of GP, for downscaling watershed precipitation in the Clutha River watershed on New Zealand's South Island.
In both studies, GP/GEP not only performed better than the SDSM, but also required fewer predictor variables.
\cite{liu2008comparison} compared the performance of a GP based method (evolutionary polynomial regression) with a feed forward neural net and the SDSM for downscaling daily total precipitation and daily maximum and minimum temperature in the Chute-du-Diable basin.
Both nonlinear machine learning methods performed about equally well and better than the SDSM, especially for precipitation.
\\
We aim at simulating the temporal evolution of spatial fields not a time series at a single grid point, i.e., we have to consider three dimensions, two in space and one in time.
We derive deterministic rules relating the fine-scale patterns to coarse model output and high-resolution information on surface properties.
Initially, we attempted to reproduce the training data fields as exactly as possible, but we quickly realized that this approach might not be appropriate for our problem.
An exact fit of the high-resolution fields is not possible, because of the inevitable unpredictable component that a downscaling procedure can not reproduce.
Furthermore, discretization, i.e., the representation of a continuous system on a discrete grid, always induces uncertainty.
However, for an improved representation of the mass and energy fluxes at the surface, modeling the sub-grid scale properties statistically might be sufficient.
A fuzzy approach for defining the fitness measure seems more appropriate than taking the exact resemblance to the high-resolution model output fields as criterion.
We thus adjusted our objective and seek to retrieve realistic spatio-temporal patterns on the fine-scale rather than reproducing the exact model output fields.
This task is very similar to the verification of spatial fields.
\\
For verification of forecasts from high-resolution numerical models fuzzy, neighborhood-based methods have already been developed during the last decades (e.g., \cite{ebert2008fuzzy}).
In the fuzzy approach it is rewarded when a model forecast is close to the observation in patterns and structures, which can be achieved by loosening the requirements for exact matches between forecast and observation comparing neighborhoods of forecast and observation points.
The Strength Pareto Approach allows the use of more than a single measure to quantify the quality of a potential downscaling rule and offers the liberty to chose the best compromise between the different objectives in hindsight.
\\
This article is structured as follows.
After introducing the data (Section 2), the downscaling methodology based on multi-objective GP is explained in detail (Section 3) and tested for the disaggregation of near-surface temperature (Section 4).  
Finally, we conclude and give an outlook on future work (Section 5).

\section{Background and Data}
\label{sec:data}
\noindent 
Our GP based downscaling is developed in order to be applied in a modeling platform, such as TerrSysMP by \cite{shrestha2014scale}, which couples land-surface and atmospheric models operating on different scales.
TerrSysMP is composed of the atmospheric model COSMO \citep{doms2002description, baldauf2011operational}, land-surface model CLM \citep{oleson2004technical} and hydrological model ParFlow \citep{kollet2006integrated}.
The three component models are coupled via the external coupler OASIS, which contains the downscaling scheme by \cite{schomburg2010downscaling}.
The scheme consists of three steps:
(1) a bi-quadratic spline interpolation interpolates the coarse-resolution atmospheric data to a higher resolution while conserving the coarse field's mean and gradients;
(2) so-called 'deterministic' downscaling rules are applied;
(3) temporal autoregressive noise is added to restore the fine-scale variance missing after step 1 and 2.
\\
The deterministic rules applied in step 2 exploit empirical relations between atmospheric variables and surface properties.
Two rules were derived from physical considerations:
near surface pressure is downscaled by exploiting the hydrostatic equation;
under cloud-free conditions, shortwave net radiation can be downscaled using surface albedo as predictor.
\\
In case there is no known physical relation the training data set was examined for possible statistical correlations between high-resolution near-surface atmospheric fields and land-surface properties.
A rule detection algorithm that calculates Pearson correlation coefficients for subsets of the training data was implemented, with the subsets selected based on indicators, e.g., cloud cover, to distinguish clear sky conditions from others.
The algorithm detected two applicable rules:
for unstable atmospheres, near-surface temperature can be downscaled using orographic height as a linear predictor;
for situations with cloud cover $< 43\%$, longwave net radiation can be disaggregated using ground temperature as linear predictor.
Under more complex conditions and for other near-surface variables (specific humidity, wind speed and precipitation) the algorithm could not find any deterministic downscaling rules applicable to a sufficiently large subset of the training data.
\\
The algorithm by \cite{schomburg2010downscaling} can only detect linear relations and is limited to dependencies between two variables.
The aim of this study is to improve the downscaling using a more flexible approach, which is able to model more complex nonlinear and multivariate relations.
This flexibility is achieved employing Genetic Programming (GP).
\\
We use the same training data set as \cite{schomburg2010downscaling}, which consists oft high-resolution (400 m) COSMO simulations on a domain of 168 km $\times$ 168 km in western Germany.
The domain is centered over the Rur catchment, which is the main investigation area of TR 32.
The data set contains hourly output for 8 simulation periods of 1 to 2 days with different weather conditions (Table \ref{table:trainingdata}).
From the simulations, we extract the inner 112 km $\times$ 112 km, i.e., 280 $\times$ 280 grid points to exclude boundary effects. 
The downscaling scheme by \cite{schomburg2010downscaling} has been trained for disaggregation from 2.8 km resolution down to 400 m and consists of three steps.
In this study we consider the same scales, i.e., we train on a downscaling by a factor of 7.

\section{Methods}
\noindent 
As motivated in the introduction, we seek to reproduce realistic spatio-temporal patterns of the near-surface atmospheric state variables, not necessarily the exact high-resolution model output.
To this goal, we formulate the downscaling problem as a multi-objective optimization problem, in order to consider multiple characteristics of the fine-scale fields, for instance the spatially distributed variance.
Summing up the different objectives to be minimized is, however, problematic since the objectives may have different units and ranges.
A scaling procedure would be required, but the risk of treating the objectives unequally or getting trapped in a local minimum would remain.
The Strength Pareto Approach (SPEA) by \cite{zitzler1999multiobjective} provides a solution to this problem.
\\
After giving a general introduction to GP, we introduce the concept of Pareto optimality SPEA is based on.
Finally, we introduce multi-objective GP step by step.
Large parts of our are based on the GPLAB package for Matlab by \cite{silva2003gplab}.
\subsection{Genetic Programming}
\noindent 
'Genetic Programming addresses the problem of automatic programming, namely, the problem of how to enable a computer to do useful things without instructing it, step by step on how to do it' (J. Koza in \cite{banzhaf1997genetic}).
GP is one of several methods within the area of evolutionary computation.
As the name implies these methods are inspired by the concept of natural evolution.
In nature an individual is exposed to environmental pressure. 
Its chance to survive, reproduce and consequently contribute to the next generation is dependent on its fitness with respect to the environmental conditions.
\\
In evolutionary computation the problem to be solved corresponds to environmental pressure.
GP evolves a solution to the problem in a similar manner as evolution happens in nature, i.e., a solution is developed over several generations, each consisting of a large number of candidate solutions.
The candidate solutions, also called individuals in analogy to the evolution terminology, are composed of program code.
\\
The candidate solutions forming the initial generation are automatically and often randomly generated and then tested on the given problem.
Each succeeding generation is evolved by applying so-called genetic operators, which recombine and modify individuals from the preceding (parent) generation.
The better a candidate solutions solves the given problem, the greater its chance is to contribute to the new generation \citep{eiben2003introduction}.
\\
Genetic Programming differs from other methods of evolutionary computing by the representation of the individuals.
In GP the individuals are traditionally represented by parse trees.
Figure \ref{fig:parsetree} shows a simple example of a parse tree consisting of 7 nodes arranged on 4 levels.
The parse tree in Figure \ref{fig:parsetree} embodies an equation consisting of arithmetic functions, variables and one constant.
A parse tree is read starting from the bottom, i.e., in figure \ref{fig:parsetree} $b$ and $1$ serve as inputs arguments to the subtraction, the output ($b-1$) and $a$ serve as input arguments to the multiplication and so on.
\\
In general a parse tree consists of functions and so-called terminals.
The set of functions and terminals used in a GP run is typically defined by the user and adapted to the problem to be solved.
A function set can contain all kinds of functions, for instance 
arithmetic functions ($+,-,\times,/$), 
transcendental functions ($\log, \sin$,...),
or conditional statements (if then else,...) \citep{banzhaf1997genetic}.
The terminals have their name, because they do not have input arguments and thus terminate the branches of the tree.
Terminals serve as fundamental input to the functions.
Thus, a terminal set can include variables, constants and zero-argument functions, such as a random number generator.
\subsection{Pareto Optimality}
\noindent 
The term Pareto optimality originates in economics.
The state of an economic system is called Pareto optimal when economic resources are distributed in such a way that it is impossible to improve the situation of one person without detereorating the situation of at least one other person.
\\
For optimization problems that involve multiple, sometimes conflicting objectives there is possibly not one single optimal solution.
Usually there exists a set of alternative solutions in which no solution is optimal in the sense that it is superior to all other solutions when considering all objectives. 
\\
The multiple objectives correspond to different quality criteria of the desired solution.
We denote the objective space containing all objective functions as $\mathcal{O}$ and the solution space containing all potential solutions as $\mathcal{Q}$.
An objective $s_i\in \mathcal{O}$ is in general calculated by comparing prediction (downscaled fields) $\textbf{y}^D$ and reference (high-resolution model output) $\textbf{y}^R$.
The prediction results from the solution (downscaling rule) $\alpha \in \mathcal{Q}$ being applied to the vector of predictors $\textbf{x}$.
Thus, incorporating all dependencies, we can write 
$s_i(\textbf{y}^D,\textbf{y}^R)=s_i(\alpha,\textbf{x},\textbf{y}^R)$.
For simplicity in the following we only include the dependency on the solution $\alpha$ explicitly.
\\
Be $\textbf{s}(\alpha)=(s_1(\alpha),s_2(\alpha),..., s_m(\alpha))^T$ the objective vector, i.e., the vector containing all $m$ objectives.
Then the multi-objective minimization problem can then be written as,
\begin{linenomath}
\begin{equation}
\textbf{s}(\alpha)=(s_1(\alpha),s_2(\alpha),...,s_m(\alpha))^T\overset{!}{=} min
\label{eq:obtprob}
\end{equation}
\end{linenomath}
Let us consider two solutions $\alpha,\beta \in \mathcal{Q}$. The solution $\alpha$ is said to dominate $\beta$ ($\alpha\succ\beta$) if and only if
\begin{linenomath}
\begin{equation}
\begin{split}
  &\forall i \in \{1,2,...,m\}:s_i(\alpha)\leq s_i(\beta)\\
   \land &\exists j \in \{1,2,...,m\}:s_j(\alpha) < s_j(\beta).
\end{split}
\end{equation}
\end{linenomath}
Or in words, $\alpha$ dominates $\beta$ if $\alpha$ is as at least as good as $\beta$ with respect to all objectives, and there exists at least one objective where $\alpha$ is better than $\beta$.
\\
The solution $\alpha$ is said to cover $\beta$ ($\alpha \succeq \beta$) if $\alpha\succ\beta$ or $\textbf{s}(\alpha)=\textbf{s}(\beta)$, i.e., either $\alpha$ dominates $\beta$ or they both perform equally well concerning all objectives.
The solutions that are not dominated by any of the elements in the solution space $\mathcal{Q}$ are called Pareto optimal.
Figure \ref{fig:speafitness} shows an example of a minimization problem with two objectives.
The squares constitute the set of Pareto optimal solutions, the circles correspond to the non-optimal solutions.
\subsection{Multi-objective Genetic Programming}
\noindent 
For multi-objective fitness assignment we have implemented the Strength Pareto Approach (SPEA) by \cite{zitzler1999multiobjective}.
SPEA requires two main changes to traditional GP.
Firstly, each generation is split into two sets, called populations $\mathcal{P}$ and $\mathcal{P}'$.
The population $\mathcal{P}$ is evolving over time as in traditional GP, whereas the second population $\mathcal{P}'$, the so-called Pareto set, contains all Pareto optimal solutions.
Secondly, the fitness calculation for individuals in both populations $\mathcal{P}$ and $\mathcal{P}'$ is based on a comparison between the individuals rather than on absolute values.
\subsubsection{Preparation}
\noindent 
GP requires a training data set to learn from, function and terminal sets to build the candidate solutions, and fitness measures (objectives) to quantify the quality of the candidate solutions.
Additionally, the user can set controlling parameters, such as population size, maximum size of the Pareto set, maximum tree size or the probability to select the various genetic operators.
Finally, a termination criterion that stops the evolution must be provided, for instance a maximum number of generations.\subsubsection{Execution}
\noindent 
Figure \ref{fig:flowchart} shows a basic flowchart of the GP algorithm incorporating SPEA, which we explain in the following.
\\
(1) An initial population of candidate solutions (individuals) is generated.
The initial population can be created randomly or include known approximate solutions of the given problem.
\\
(2) Each candidate solution (each potential downscaling rule) is applied to the training data set.
\\
(3) The result of each candidate solution is evaluated according to the objectives.
\\
(4) The Pareto set is updated:
All individuals in population $\mathcal{P}$ that are nondmiated within $\mathcal{P}$ are moved to the Pareto set $\mathcal{P}'$.
The individuals in $\mathcal{P}'$ that are covered by another member of $\mathcal{P}'$ are removed from the Pareto set.
In case the number of individuals stored in $\mathcal{P}'$ exceeds the given maximum, $\mathcal{P}'$ is pruned by hierarchical clustering.
A clustering procedure helps to preserve the solution diversity while shrinking the Pareto set.
To make sure that all objectives are considered equally, we scale the values before applying the clustering procedure.
As in our case, each objective $s_i$ is positively defined, we calculate the scaled objective as $s_i^{sc}(\alpha)=\Big( s_i(\alpha)- \underset{\beta \in P'}{min}(s_i(\beta)) \Big)/\underset{\gamma \in P'}{max}(s_i(\gamma))$, i.e., from the objective we subtract the minimum and divide the result by the maximum occurring in the current Pareto set.
The fitness results from a greater than/smaller than comparison, i.e., scaling does not effect the fitness.
Note that the scaling is only applied within the clustering step.
\\
(5) The fitness of each individual in $\mathcal{P}$ and $\mathcal{P}'$ is calculated according to SPEA by comparing the individual's performances.
(For details see Section \ref{sec:SPEAfit})
\\
(6) If the termination criterion is met, the final Pareto set is returned.
If the termination criterion is not met, the algorithm continues with (7).
\\
(7) The next generation is created by combining and mutating individuals from the current $\mathcal{P}+\mathcal{P}'$.
The creation of the new generation can be split into two steps.
First, a sampling procedure is applied to determine the parents.
Second, genetic operators (crossover, mutation) are applied to create new individuals.
\\
(7.1) For sampling we use the lexicographic parsimony pressure \citep{luke2002lexicographic} as it is implemented in GPLAB.
A number of individuals is randomly drawn from the current $\mathcal{P}+\mathcal{P}'$.
The individual drawn with the best fitness is to become parent.
In case several individuals are equally fit the smallest one is chosen.
\\
(7.2) The genetic operators are applied to the individuals.
Crossover recombines two parents.
The parent parse trees are cut at a randomly chosen node and the separated subtrees are exchanged.
(Subtree)-mutation cuts a randomly chosen subtree from the parent and replaces it by a new randomly created subtree.
Parent selection and application of genetic operators, are repeated until enough individuals for the new generation have been created.
\\
Starting from (2) the succeeding steps are iteratively repeated until the termination criterion is met (see (6)).
\subsubsection{Fitness Assignment}
\label{sec:SPEAfit}
\noindent 
In SPEA the fitness assignment consists of two steps.
\\
(1) Each solution in the Pareto set, $\alpha\in P'$, is assigned a real value called fitness $f'(\alpha) \in [0,1)$.
The fitness $f'(\alpha)$ is proportional to the number of individuals $\beta \in P$ that are covered by $\alpha$, i.e., $\alpha \succeq \beta$.
Let $N$ be the total number of individuals in $\mathcal{P}$.
Then $f'(\alpha)$ is defined as $f'(\alpha)=\frac{n(\beta|\alpha \succeq \beta, \alpha\in P',\beta \in P)}{N+1}$.
To clearly separate between the fitness of individuals in $\mathcal{P}$ and $\mathcal{P}'$, the fitness of the individuals in $\mathcal{P}'$ is also called strength, hence Strength Pareto Approach.
\\
(2) The fitness $f(\beta)$ of an individual in the population $\mathcal{P}$, $\beta \in P$, is calculated as the sum over the fitness of all individuals in the Pareto set, $\alpha \in \mathcal{P}'$, that cover $\beta$,
$f(\beta)=1+\sum_{\alpha,\alpha\succeq \beta}f'(\alpha)$, where $f(\beta) \in [1,N)$.
One is added to the sum to ensure the individuals in the Pareto set $\mathcal{P}'$ have better fitness than those in $\mathcal{P}$.
\\
Figure \ref{fig:speafitness} shows one possible scenario of a minimization problem with two objectives $s_1$ and $s_2$.
The values indicate the fitness.
The circles correspond to the individuals in $\mathcal{P}$, the squares to the individuals in $\mathcal{P}'$.
The lowest point in Figure \ref{fig:speafitness} shows an individual contained in the Pareto set $\mathcal{P}'$ that dominates 3 out of the 7 individuals in $\mathcal{P}$.
Therefore its fitness $f'$ equals $3/(7+1)=3/8$.
The next lowest point represents an individual from the population $\mathcal{P}$, which is dominated only by one individual with a fitness $f'$ of $3/8$.
Hence its fitness $f$ calculates as $1+3/8=11/8$.
\subsection{Objectives}
\noindent 
Our downscaling aims at reproducing the fine-scale patterns, i.e., reproducing as much of the deterministic part as possible, while accounting for spatial noise or uncertainty. 
Combining both aspects makes the original and the downscaled fields 'look similar'.
Note that we aim at predicting anomalies, i.e., the difference between the high-resolution and the spline-interpolated fields.
In the following this is not always explicitly mentioned.
\\
One possibility to make two fields look similar is to make their values similar in distribution. 
Thus our first objective is to minimize the differences in distribution between the downscaled and the reference values (anomalies) thereby completely ignoring the location of the values.
This difference is measured by the integrated quadratic distance $IQD$.
The $IQD$ quantifies the distance between two cumulative distribution functions ($CDFs$) $F$ and $G$ \citep{thorarinsdottir2013using}, 
\begin{linenomath}
\begin{equation}
IQD=\int_{-\infty}^{\infty}(F(x)-G(x))^2 dx.
\end{equation}
\end{linenomath}
We apply the $IQD$ to the discrete histogram distributions.
The histogram distributions are calculated using a bin width of 0.25 K.
Let $H(y_t^R)$ be the normalized histogram distribution for the high-resolution reference field at time step $t$, and accordingly $H(y_t^D(t))$ the normalized histogram distribution for corresponding the downscaled field.
A single histogram bin is denoted by $H_i$.
From the histogram distributions, we calculate $IQD$ for each field (time step) separately (Figure \ref{fig:scales1}) and take the mean over all time steps as objective, i.e.,
\begin{linenomath}
\begin{equation}
IQD=1/n_t\sum_t\sum_i(H_i(y_t^R))-H_i(y_t^D))^2,
\end{equation}
\end{linenomath}
where $n_t$ denotes the number of fields contained in the reference data set.
\\
The deterministic part is accounted for  by using the root mean square error ($RMSE$) as objective.
The $RMSE$ provides an estimate of the expected value, and smooths out those patterns of the fine scale fields that are not deterministically reproducible.
To account for some uncertainty we introduce a fuzzy method.
In our application we account for uncertainty in the location of the fine-scale structures by defining a neighborhood 
$U(i,j)=\{(i,j),(i,j+1),(i,j-1),(i+1,j),(i-1,j)\}$ for each pixel $(i,j)$.
This neighborhood contains the four direct neighbors of $(i,j)$ (Figure \ref{fig:scales1}).
We incorporate the neighborhood in the calculation of the $RMSE$ as follows:
\begin{linenomath}
\begin{equation}
RMSE=\sqrt{ \frac{1}{n_t n_i n_j} \sum_{i,j,t} \underset{k,l \in U(i,j)}{min} (y_{tij}^R - y_{tkl}^D) ^2 }.
\end{equation}
\end{linenomath}
Here, $n_i$ and $n_j$ denote the total number of pixels in x- and y-direction.
\\
As noted, the spatial variability of the non-deterministic part is smoothed out by minimizing the $RMSE$. 
In order to recover variability, we use the mean error of the coarse pixel standard deviation $ME(STD)$ as third objective.
Let $V(p,q)$ denote a pixel on the coarse scale containing $7 \times 7$ pixels on the finer scale (Figure \ref{fig:scales1}).
Note that unlike $U(i,j)$, $V(p,q)$ is not defined as a sliding neighborhood, but via a fixed grid, i.e., the grid of the coarse model output to be downscaled, due to computational reasons.
\\
The coarse pixel standard deviation $\sigma(y^R_{tpq})$ of the reference field $t$ calculates as 
\begin{linenomath}
\begin{equation}
\sigma(y^R_{tpq}) =\sqrt{\frac{1}{7 \times 7 -1} \underset{i,j \in V(p,q) }{\sum}(y_{tij}^R-\bar{y}_{tpq}^R)^2},
\end{equation}
\end{linenomath}
with $\bar{y}_{tpq}^R$ denoting the coarse pixel mean. 
Since we are dealing with anomalies $\bar{y}_{tpq}^R=0$.
We can now define the third objective as
\begin{linenomath}
\begin{equation}
ME(STD)=\frac{7 \times 7}{n_i n_j n_t}\sum_{t,p,q}\Big|\sigma(y^R_{tpq})-\sigma(y^D_{tpq})\Big|.
\end{equation}
\end{linenomath}
The size of the solution's parse trees, i.e., the number of nodes, serves as the fourth objective.
Smaller parse trees can be checked for physical consistency more easily and are computationally less expensive.
Incorporating the size as objective gives furthermore information on a potential dependency between the quality and the complexity of the solutions.
\\
In summery, we have three objectives quantifying the quality of the potential downscaling rules by considering different characteristics of the fine-scale fields at different scales, and one objective incorporating the solution's complexity.

\section{Downscaling near-surface temperature}
\noindent 
We apply multi-objective GP to downscaling of near-surface temperature at 10 m height, which is the center of the lowest atmospheric layer in the COSMO model.
High-resolution temperature fields can exhibit very different fine-scale patterns.
In a well mixed ABL, temperature anomalies are in good approximation proportional to orographic height anomalies \citep{schomburg2010downscaling}.
Under clear sky conditions during night time, temperature inversions cause cold air to drain into the valleys  \citep{barr1989influence}, which leads to pronounced channel structures in the temperature field.
The anomalies caused by cold air drainage can grow very large compared to the anomalies in a well-mixed boundary layer (Figure \ref{fig:dc_rmse}).
Therefore, it is important to also capture this more complex process in the downscaling algorithm.

\subsection{Setup}
\noindent 
To test multi-objective GP for atmospheric downscaling, we use a cross-validation approach (i.e., leave one out) for three reasons.
First, it allows us to use a training data set that is as big as possible in each GP run.
Second, it allows us to check for overfitting, a well known problem especially when training on a small data set.
With a cross-validation approach, we can identify weather conditions for which our solutions can not be generalized.
Third, we gain information on the impact of the training data choice on the solutions by considering the similarity of GP solutions resulting from different training data sets.
\\
As mentioned in Section 2 the complete data set contains 8 simulation periods (Table \ref{table:trainingdata}).
For computational reasons we have extracted single time steps for training and validating GP.
We have carried out 8 GP runs in total.
In each run another day is left out for validation.
\\
The potential predictors included in the terminal set are preselected from the COSMO model output based on our understanding of atmospheric processes, which influence near-surface temperature.
The coarse-scale fields we pick are near-surface temperature, vertical temperature gradients of the lowest 25 m, 60 m, and 110 m as indicators for atmospheric stability, horizontal and vertical wind speed, and net radiation at the land surface.
Fine-scale information is contained in the parameters describing the surface properties.
We pick topographic height, plant cover, surface roughness length and surface albedo, as well as a few parameters derived from the topography field.
The latter give information on local topography relative to its direct surroundings.
All predictors are listed in Table \ref{table:predictors}.
\\
The essential GP settings are summarized in Table \ref{table:GPsettings}. 
We run 200 generations with 100 individuals each, i.e., each run evaluates 20000 potential downscaling rules.
The maximum Pareto set size is set to 50.
For computational reasons and to keep the solutions readable, we furthermore limit the tree size to 5 levels.
Besides predictors described above the terminal set contains also a random number generator.
The function set contains the arithmetic functions with 2 input argument each and an if-statement with four input arguments (i.e., if $a>b$ do $c$ else do $d$).

\subsection{Results}
\noindent 
Figure \ref{fig:scatterPareto} shows the relative reduction of objectives for the solutions from the Pareto sets returned by the 8 GP runs.
Only the first three objectives, $RMSE$, $ME(STD)$ and $IQD$, are included in Figure \ref{fig:scatterPareto}.
These objectives are formulated as penalties, i.e., the smaller the objective the better.
We calculate the relative improvement for a downscaling rule $\alpha$ concerning an objective $s_i$ as, $\tilde{s_i}(\alpha)=1-s_i(\alpha)/s_i(0)$,
where $s_i(0)$ is the objective when predicting no anomalies, i.e., zero everywhere.
A prediction of no anomalies corresponds to the spline-interpolated field.
The definition of the relative improvement is analogue to that of a skill score, with the objective for a perfect solution being equal to zero.
A positive $\tilde{s_i}(\alpha)$ indicates that the downscaled field is better than the spline-interpolated field concerning objective $s_i$; for a perfect downscaling $\tilde{s_i}(\alpha)=1$; when the objective is halved $\tilde{s_i}(\alpha)=0.5$; for a downscaling that is as good as the interpolated field $\tilde{s_i}(\alpha)=0$; for a downscaling worse than the interpolated field $\tilde{s_i}(\alpha)<0$.
Note that $\tilde{s_i}(\alpha)\in(-\infty,1]$, i.e., in the positive direction $\tilde{s_i}(\alpha)$ can not exceed 1, whereas it can grow very large in the negative direction.
\\
Looking at Figure \ref{fig:scatterPareto}, we see that for the three different objectives, we achieve very different improvements.
The relative improvement of the $ME(STD)$ spans a wide range from 0 up to about 0.7 with an average of about 0.4-0.5.
For the $IQD$ the relative improvement is in general smaller with an average of about 0.1-0.2.
Unlike for the $ME(STD)$, for the $IQD$ there are also solutions which perform worse than the interpolated field.
The $RMSE$ is on average even increased by approximately $10 \%$.
Only very few solutions achieve an improvement concerning the $RMSE$.
This shows the difficulty of predicting the actual reference fields exactly.
Including the $RMSE$ as objective, however, serves as a constraint not to deviate too strongly from the reference fields.
The results confirm that predicting realistic spatial variability is better attainable than an exact fit.
\\
To study potential overfitting, Figure \ref{fig:boxplot} shows the difference of the relative improvement between training and validation data set ($\tilde{s}_{tr}-\tilde{s}_{val}$).
For the majority of cases the median is close to zero.
With the exception of May 9th 2008, the medians are spread about equally into positive and negative direction, which indicates that no serious overfitting takes place and that the downscaling rule discovered by GP are adaptable.
For most of the 8 cases there are very few outliers ($\approx$ 2-6 out of 50) for which the performance on the validation data set is clearly worse compared to the training data set.
The outliers are most apparent for the $RMSE$.
Most outliers show already a less satisfying performance on the training data sets and tend to stick out in the scatter plots (e.g., Figure \ref{fig:scatterPareto} (a) or (c)).
These solution are part of the Pareto set due to their good perform in the 4th objective, i.e., the solution size.
They correspond to very small solutions and often consist of only one node (Figure \ref{fig:scatterqual}).
We seek for a reasonable tradeoff between the objectives and hence also between quality and complexity of the solutions.
Thus, such extreme outliers would not be chosen from the Pareto set as the final solution.
\\
Overfitting could be a problem for the case where May 9th 2008 is excluded for training.
The clear sky conditions on May 9th 2008 led to very pronounced fine-scale structures.
Probably, the exclusion of such an extreme situation led to the bad performance for this case.
For this case the solutions need to extrapolate and thus perform in general worse on the validation day than on the training data set.
\\
Besides the discussed outliers, the solution quality is only slightly dependent on the solution size.
Figure \ref{fig:scatterqual} shows scatter plots for all four objectives for the run validated on October 14th 2007.
The solution size corresponds to the abscissa in Figure \ref{fig:scatterqual} (d)-(f).
There is only a week correlation between solution size and the other objectives. 
The subjective rating of the solutions indicated by the color of the points in Figure \ref{fig:scatterqual} is explained in the discussion.
\\
We examine the results of a single solution from the run validated on October 14th 2007, a clear sky day.
Figure \ref{fig:fieldday} and \ref{fig:fieldnight} show the performance one downscaling rule on the validation day, at 11:00 UTC and 23:00 UTC.
The daytime temperature field (Figure \ref{fig:fieldday}) is well predicted by both the GP based rule as well as by the (thresholded) linear regression based rule from \cite{schomburg2010downscaling}.
However looking more closely and zooming in, the fine-scale structures are seemingly better captured by the GP based rule.
For the nighttime field (Figure \ref{fig:fieldnight}) the improvement by GP more impressive:
the pattern formed by nightly cold air drainage is much better captured by the GP based rule.
We notice that the minima and maxima are localized more strongly in the reference field than in the downscaled field, which might be at least partly due to the unpredictable noisy component of the fine-scale fields.
\\
The GP derived rule shown in Figures \ref{fig:fieldday} and \ref{fig:fieldnight} is:
\\
\\
if $HSURF_a > 0.98$\\
then $HSURF_a \times T_{gr60}$\\
else $HSURF_a \times T_{gr110}$\\
+ if $T_{gr25} > T_{gr110}$\\
\noindent\hspace*{3mm} then  if $0.83 > T_{gr25}$\\%
\noindent\hspace*{11.2mm} then $w_v$\\
\noindent\hspace*{11.2mm} else $Topo_2$\\
\noindent\hspace*{3mm}  else $w_v$
\\
\\
In the first part, surface height anomaly $HSURF_a$ is multiplied by the temperature gradients $T_{gr}$, which is physically intuitive and reasonable.
The second part either adds vertical wind speed $w_v$ or one of the topography based parameters $Topo_2$ depending on the atmospheric layering.
\\
Note that the example solution shows some evidence for overfitting.
The relative improvements of the objectives for training data set and validation day are listed in Table \ref{table:objexample}.
Still the performance on the validation day appears reasonable, since no irrealistically high or low values occur.
\subsection{Discussion}
\noindent 
Multi-objective GP has advanced downscaling in two directions compared to linear regression approach by \cite{schomburg2010downscaling, schomburg2012disaggregation}.
Firstly, it allows to incorporate multivariate and nonlinear relations in the search for downscaling rules and secondly, uncertainties contained in the training data set are considered by defining multiple objectives.  
Nonlinear regression via rather complex equations or code can induce overfitting, especially when applied to small training data sets.
In the GP based methodology the tendency to overfit is rather small.
It is, however, important to span a wide range of weather conditions including extreme days to prevent overfitting.
\\
Incorporating SPEA, GP does not return one single downscaling rule but 50 of quite different quality when visually comparing downscaled and reference fields.
The colors of the points in Figure \ref{fig:scatterqual} indicate the results of a subjective evaluation of the different solutions by examination of the downscaled fields.
Good, average and bad solutions can obviously not be clustered easily, but some systematics are visible.
Figure \ref{fig:scatterqual} (d) indicates that a certain level of $RMSE$ must be accepted in order to reasonably reproduce the fine-scale variability, because subjectively good solutions tend to perform worse concerning the $RMSE$ than visually bad solutions.
This prompts the question, if the $RMSE$ is an appropriate objective for our aim, or if other measures are better suited.
$ME(STD)$ appears to be a more important objective.
The better a solution concerning the $ME(STD)$, the better it is also in the subjective evaluation.
Concerning the $IQD$ the situation is less clear, but a similar tendency exists as for the $ME(STD)$.
When subjectively rating the solutions, more complex (larger) solutions tend to perform better.
Thus, a minimum solution size is required to account for the complexity of the processes involved in building the fine-scale structures.
However, the best solution seem not to be detectable from the objectives alone.
Figures \ref{fig:fieldday2} show a solution we visually rate as average, a conclusion we could not draw from the objectives alone.
Only looking at the objectives the two solutions seem very similar.
Thus it might be necessary to reconsider the objectives we currently use, to find a better way of quantifying the similarity between the fine-scale patterns of downscaled and reference fields 
\\
We tested the errors of temporal and spatial correlation as additional objectives.
However, this caused the solution size to increase considerably.
Temporal correlation between consecutive time steps, as well as nearest neighbor correlation might be too strongly dependent on the noisy component of the fields and therefore might be unpredictable deterministically.
The coarse scale predictors tend to be rather smooth and thus the correlation between consecutive time steps tends to be overestimated.
Reproduction of spatio-temporal correlations might be more appropriately achieved by adding spatial and temporally correlated noise to the deterministic fields.

\section{Conclusion and Outlook}
\noindent 
We have introduced multi-objective GP for the discovery of downscaling rules to disaggregate meso-scale atmospheric model output to the high resolution required for driving land-surface and hydrological models.
Applying multi-objective GP to downscaling near-surface temperature, we have shown that the more complex method is an advancement compared to linear regression conditioned on indicators.
We are now capable to reproduce fine-scale patterns resulting from different atmospheric processes, e.g., produced in a well mixed boundary layer or during nighttime temperature inversions.
In general we can recommend GP as a good machine learning method for the geosciences, especially for complex problems incorporating multiple objectives where linear methods do not give satisfactory results.
\\
Next steps will include the optimization of the set of objectives, as discussed, and the expansion of the training data set, especially to include more clear-sky days and extreme weather conditions.
We will apply multi-objective GP to the remaining variables required as atmospheric input to land-surface and subsurface models, i.e., precipitation, near-surface specific humidity, near-surface wind speed, long-wave and short-wave radiation.
The discovered downscaling rules will be implemented in the coupled soil-vegetation-atmosphere model TerrSysMP.
Finally we will extend the system for different scale gaps, e.g., for disaggregating from 1 km to 100 m.
\\
Multi-objective GP generates a set of Pareto optimal solution and not just one, which suggests the implementation of a downscaling ensemble by either using different rules in each ensemble run or by randomly switching between different rules taking into account temporal consistency.
An ensemble approach could also help to estimate the source of uncertainty induced by the downscaling procedure.


\section*{Acknowledgments}
\noindent 
This work has been carried out in the framework of the Transregional Collaborative Research Centre TR 32 'Patterns in Soil-Vegetation-Atmosphere-Systems' funded by the Deutsche Forschungsgemeinschaft (DFG).

\bibliographystyle{agsm}


\bibliography{thebib}

\clearpage
\input tablesX.tex

\clearpage
\input figuresX.tex

\end{document}

%% file: tablesX.tex
\begin{table*}
\caption{Predictors included in the terminal set. The fields of the atmospheric state variables are given at coarse resolution (i.e., 2.8 km), the constant surface property fields are given at high resolution (i.e., 400 m).}
\label{table:predictors}
\centering
\begin{tabular}{l l}
\hline
\hline
\textbf{Weather information (coarse)}\\
\hline
$T$ & near surface temperature \\
$T_{gr25}$ & vert. temp. gradient of lowest 2 layers ($\approx 25 m$)\\
$T_{gr60}$ & vert. temp. gradient of lowest 3 layers ($\approx 60 m$) \\
$T_{gr110}$ & vert. temp. gradient of lowest 4 layers ($\approx 110 m$)\\
$w_v$ & near surface vertical windspeed \\
$w_h$ & near surface horizontal windspeed \\
$R_{net}$ & net radiation \\
\hline
\hline
\textbf{Surface information (high-res.)}\\
\hline
$HSURF_a$ & topographic height anomaly\\ 	
$Topo_1$ & mean height difference to neighboring grid points \\
$Topo_{1a}$ & anomaly of $Topo_1$ \\
$Topo_2$ & slope to lowest neighboring grid point \\
$Topo_3$ & slope to highest neighboring grid point \\
$Topo_4$ & number of direct neighbors lower than grid point\\
$PLC$ & plant cover \\
$RH$ & roughness length	\\
$ALB$ & albedo \\
\hline
\hline
\end{tabular}
\end{table*}

\begin{table*}
 \caption{Simulation dates and weather situations used for training and validation.
 To weight all simulation periods equally, we only include one day of each simulation period in the GP training data sets.
 The day in brackets is not used.
 The right column lists the time steps we have extracted for training GP due to computational reasons.}
 \label{table:trainingdata}
\begin{tabular}{l l l}
\hline
\hline
\textbf{Date}   & \textbf{Weather} &\textbf{Time steps used in GP runs} \\
\hline

27 Aug. 2007 	&	 varying cloud cover, no precipitation & 3:00-4:00, 15:00-16:00\\

14,(15) Oct. 2007 &	 clear sky & 11:00-12:00, 23:00-24:00\\

10 March 2008   &	 strong winds, variable clouds and precipitation & 10:00-11:00, 22:00-23:00 \\

(1),2 May 2008 	& 	 clouds and precipitation & 0:00-1:00, 12:00-13:00\\

(9),10 May 2008	&	 clear sky & 1:00-2:00, 13:00-14:00\\

7,(8) June 2008	&	 convective clouds and precipitation & 5:00-6:00, 17:00-18:00\\

21 July 2008	&	 synoptically driven stratiform rainfall & 9:00-10:00, 21:00-22:00\\

28 Aug. 2008	&	 cloudy, some rain & 7:00-8:00, 19:00-20:00 \\
\hline
\hline
\end{tabular}
\end{table*}

\begin{table*}
\caption{Summary of the GP settings. (Protected division means that division by zero returns the dividend not an error.)}
\label{table:GPsettings}
\centering
\begin{tabular}{l l}
\hline
\hline
\textbf{Parameter}	& 	\textbf{Value}	 \\
\hline
function set 		&	 +,-,*,protected /, if	\\

terminal set		& 	random numbers [0,1], variables (Table \ref{table:predictors})\\

generations		&	200 	\\

population size		&	100	\\

max. Pareto set size	&	50	\\

genetic operators	&	(subtree-)mutation, crossover	\\


max. tree size		&	5 levels \\
\hline
\hline
\end{tabular}
\end{table*}

\begin{table*}
 \caption{Relative reduction of objectives for the downscaling rule shown in Figures \ref{fig:fieldday} and \ref{fig:fieldnight}.}
\label{table:objexample}
\begin{center}
\begin{tabular}{c|ccc}
  & RMSE & ME(STD) & QD\\
\hline
training & -0.18 & 0.53 & 0.21\\
validation & -0.49  & 0.56 & 0.09
\end{tabular}
\end{center}
\end{table*}

%% file: figuresX.tex
\begin{figure*}
\centering
     \includegraphics[width=0.15\textwidth]{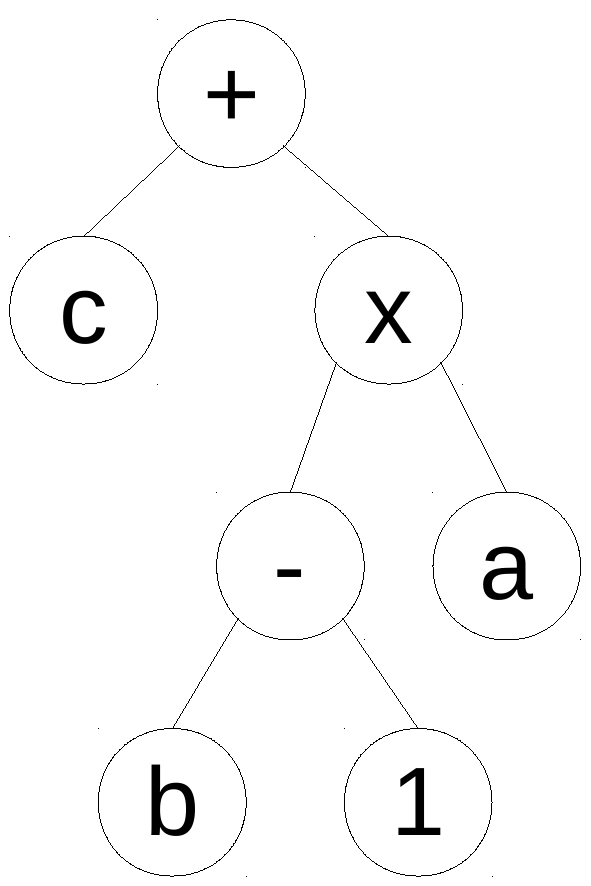}
	\caption[]{Example of a parse tree representing the equation $a \times (b-1)+c$.}
\label{fig:parsetree}
\end{figure*}

\begin{figure*}
\centering
     \includegraphics[width=0.4\textwidth, trim=0cm 2cm 0 0cm]{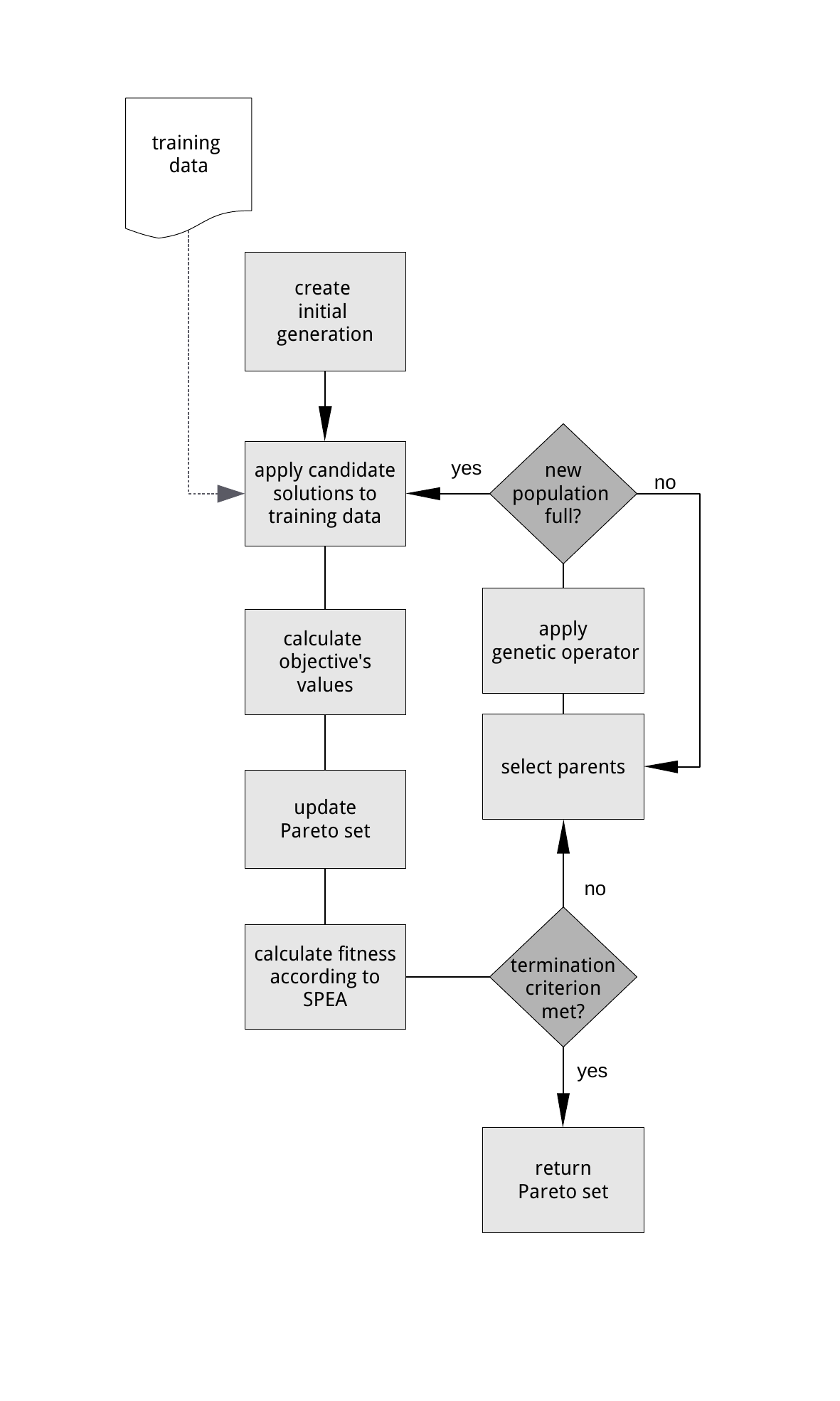}
	\caption[]{Flowchart showing the essential steps of Genetic Programming with multi-objective fitness assignment according to the Strength Pareto Approach by \cite{zitzler1999multiobjective}.}
\label{fig:flowchart}
\end{figure*}

\begin{figure*}
\centering
     \includegraphics[width=0.4\textwidth]{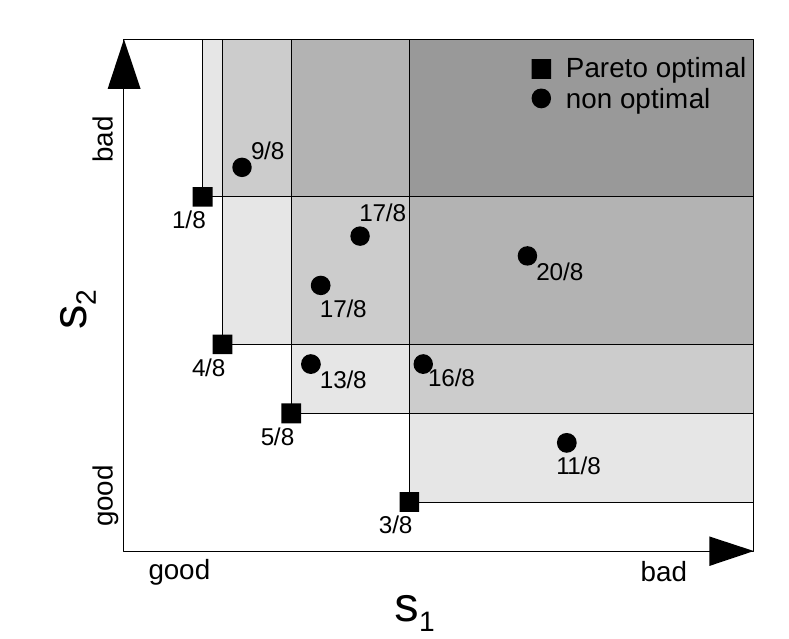}
	\caption[]{Example of a minimization problem with two objective ($s_1$ and $s_2$).
The squares correspond to the Pareto optimal solutions; the circles to the non optimal solutions.
The number associated with each solution gives the fitness according to the Strength Pareto Approach.
The figure is adapted from \cite{zitzler1999multiobjective}.}
\label{fig:speafitness}
\end{figure*}

\begin{figure*}
\centering
     \includegraphics[width=0.4\textwidth, trim=0cm 1cm 0 0cm]{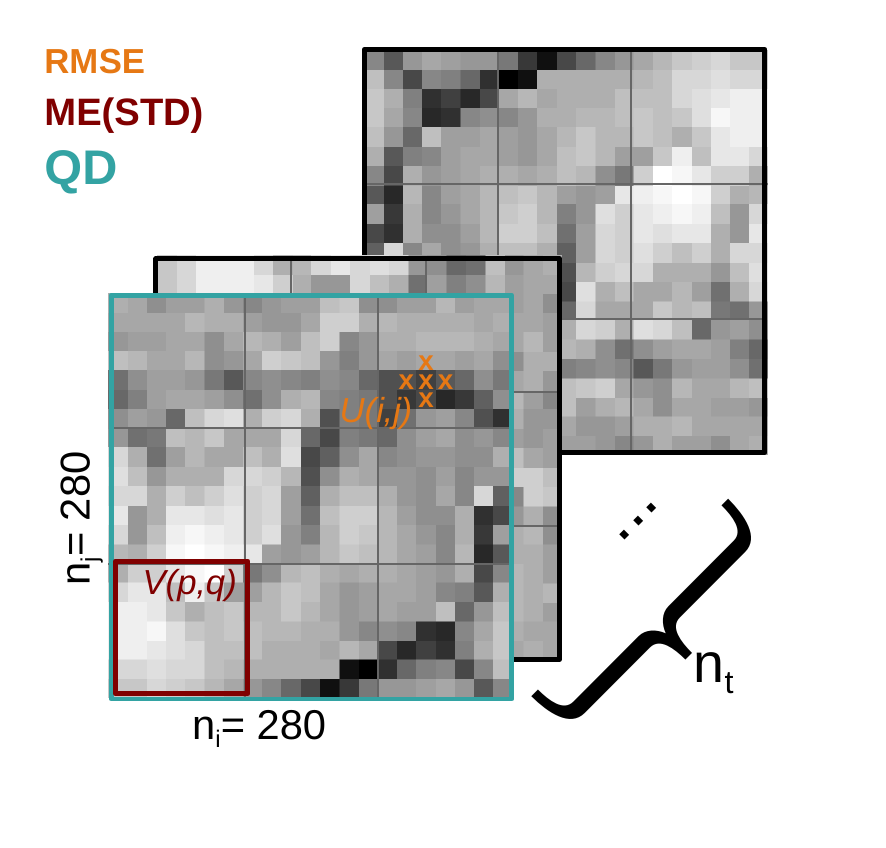}
	\caption[]{Sketch of the scales the where different objectives are defined:
the quadratic distance $IQD$ is calculated from the full fields;
the mean error of standard deviation $ME(STD)$ is defined on the coarse pixels $V(p,q)$;
the $RMSE$ is defined on the fine-scale pixels $(i,j)$ incorporating the pixel's neighborhood $U(i,j)$;
$n_i$, $n_j$, and $n_t$ are the training data dimensions.}
\label{fig:scales1}
\end{figure*}

\begin{figure*}
\centering
     \includegraphics[width=0.45\textwidth]{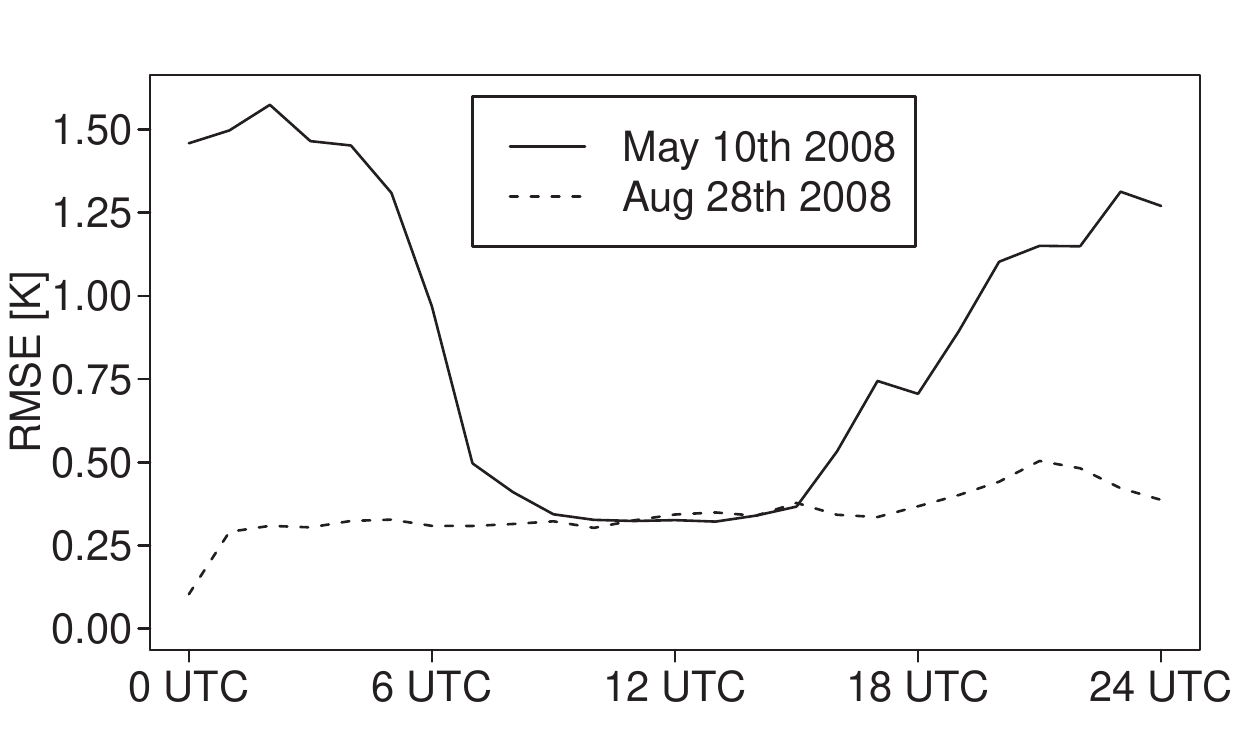}
	\caption[]{Diurnal cycle of the root mean square error of the spline-interpolated fields w.r.t. the high-resolution reference fields.
		   May 10th 2008 was an almost clear sky day.
		   August 28th 2008 was a cloudy day with some rain.}
\label{fig:dc_rmse}
\end{figure*}

\begin{figure*}
\centering
	\includegraphics[width=0.8\textwidth, trim=0cm 0cm 0 0cm]{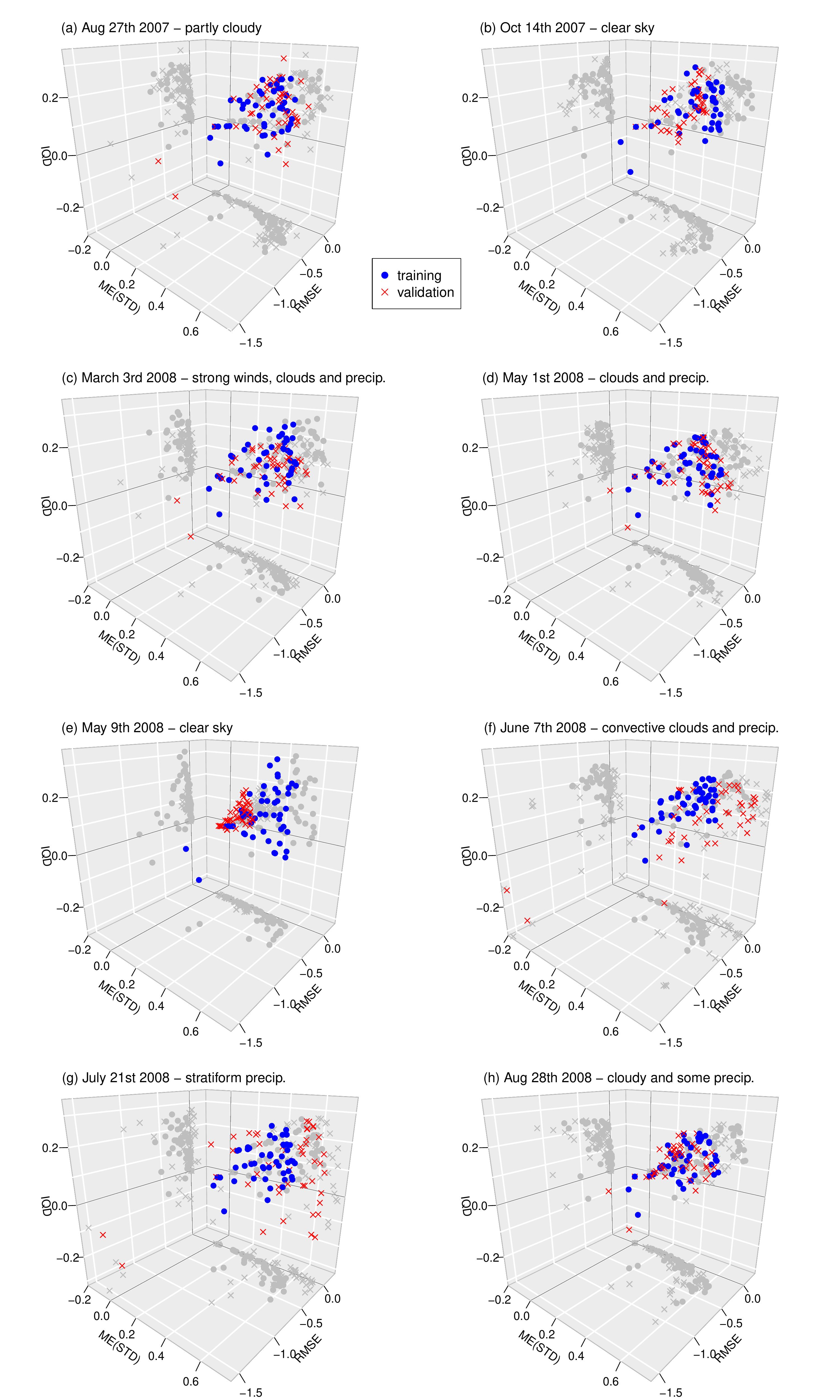}
	\caption[]{Relative reduction of objectives.
		   The 8 subfigures correspond to the 8 GP runs, each containing 50 blue circles indicating the performance of each solution of the returned Pareto set on the training data and 50 red crosses indicating the performance on the validation day:
		   (a) shows the results validated on August 27th 2007, and so on.}
\label{fig:scatterPareto}
\end{figure*}

\begin{figure*}
\centering
	
	\includegraphics[width=0.8\textwidth]{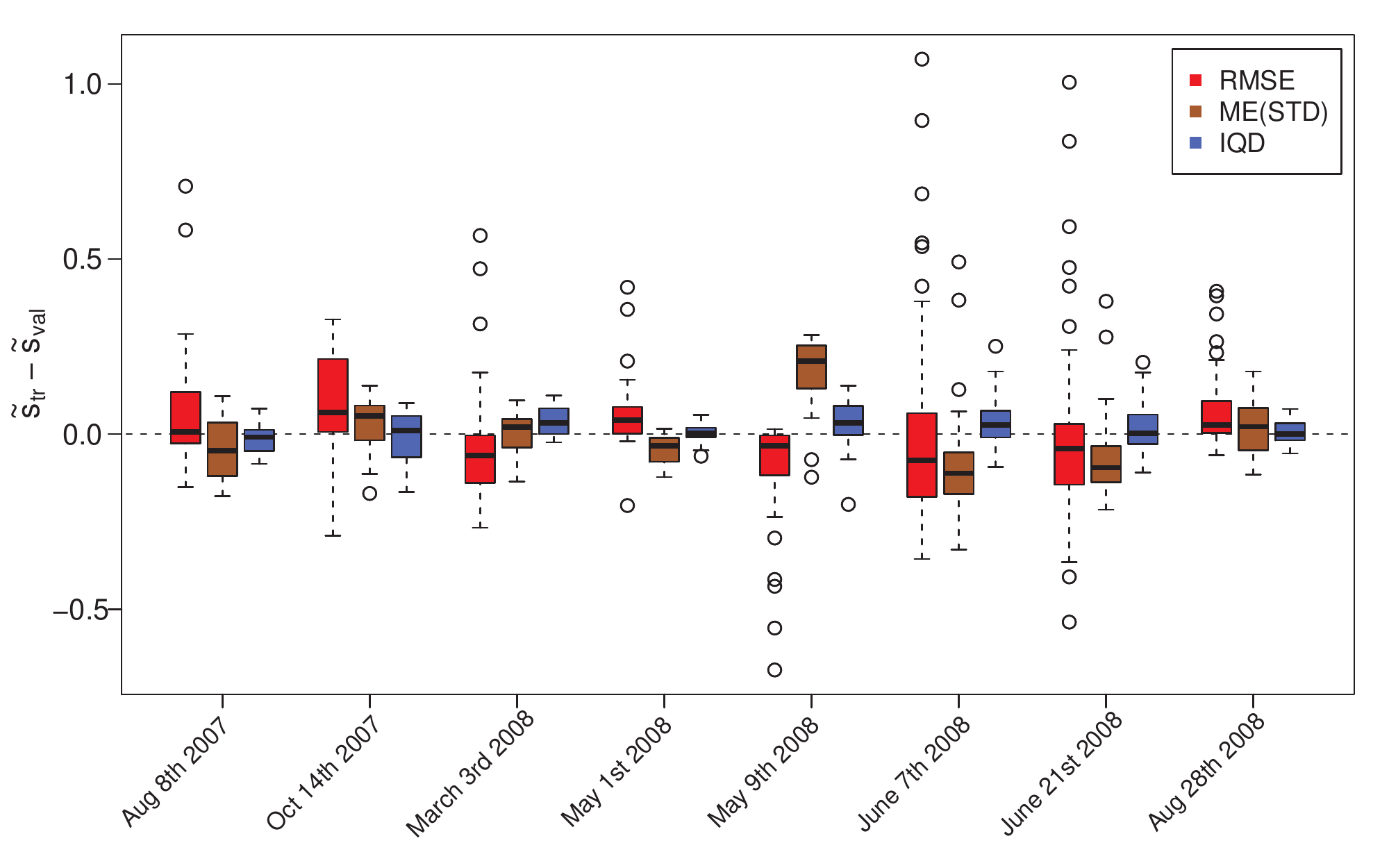}
	\caption[]{Relative reduction of objectives: difference between training and validation data set ($\tilde{s}_{tr}-\tilde{s}_{val}$) for all 8 runs.
		   Each box results from 50 values, one for each solution from the Pareto set.
		   The horizontal line within the boxes is the median, the upper and lower boundaries of the boxes correspond to the $75\%$- and $25\%$-quantiles.
		   The whiskers indicate the range spanned by maximum and minimum.
		   The length of the whiskers is however restricted to $1.5\times$box size.
		   Values outside this range are considered outliers and are shown as circles.
		   }
\label{fig:boxplot}
\end{figure*}

\begin{figure*}
\centering
	\includegraphics[width=\textwidth]{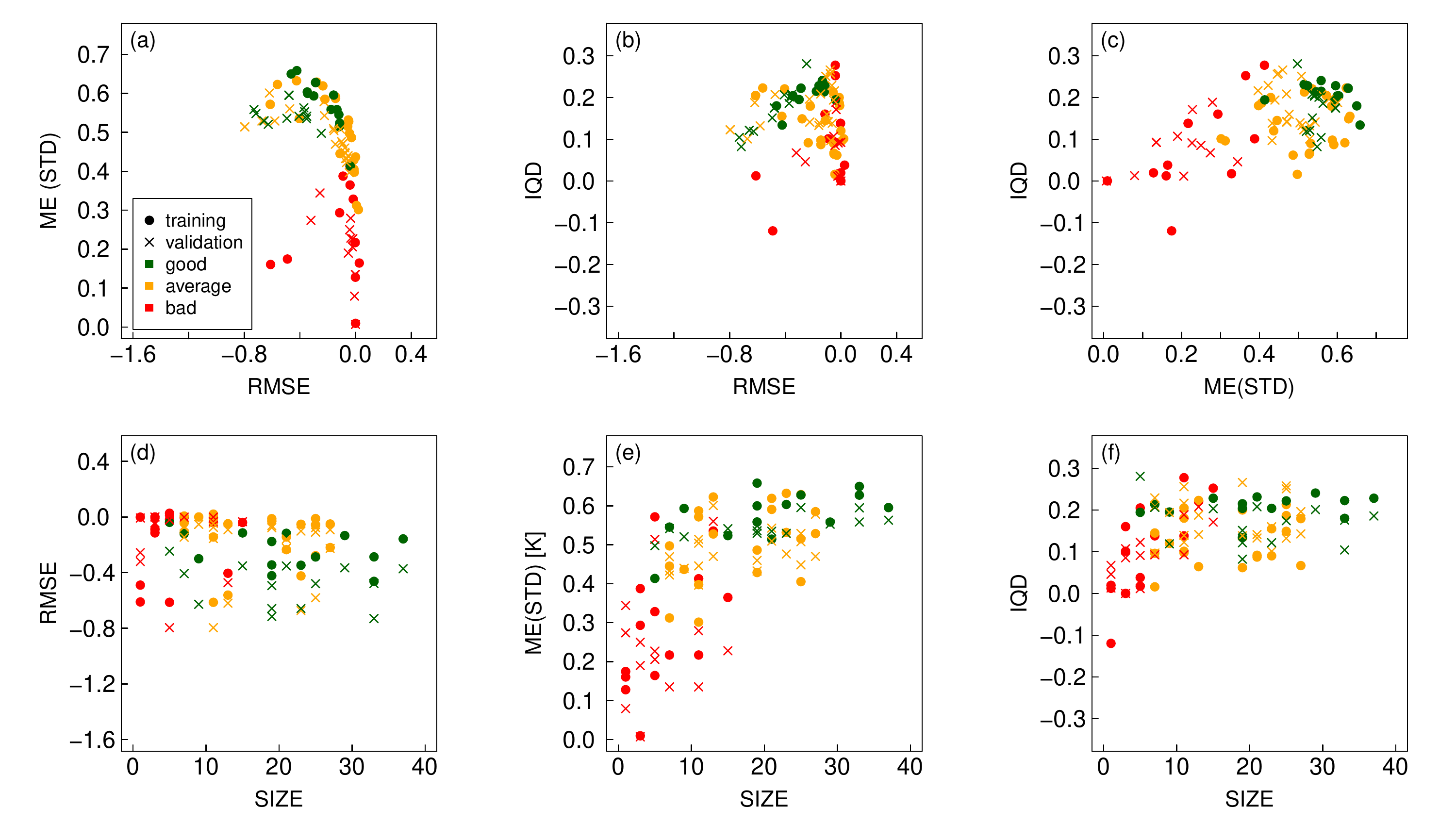}
	\caption[]{Scatterplots incorporating all four objectives: relative reduction of objectives for $RMSE$, $ME(STD)$ and $IQD$; absolute value for $SIZE$. 
		   Shown are the results validated on October 14th 2007.        
		   The colors indicate how we subjectively rate the performance when visually comparing downscaled and reference fields on October 14th 2007.}
\label{fig:scatterqual}
\end{figure*}

\begin{figure*}
\centering
	\includegraphics[width=0.9\textwidth]{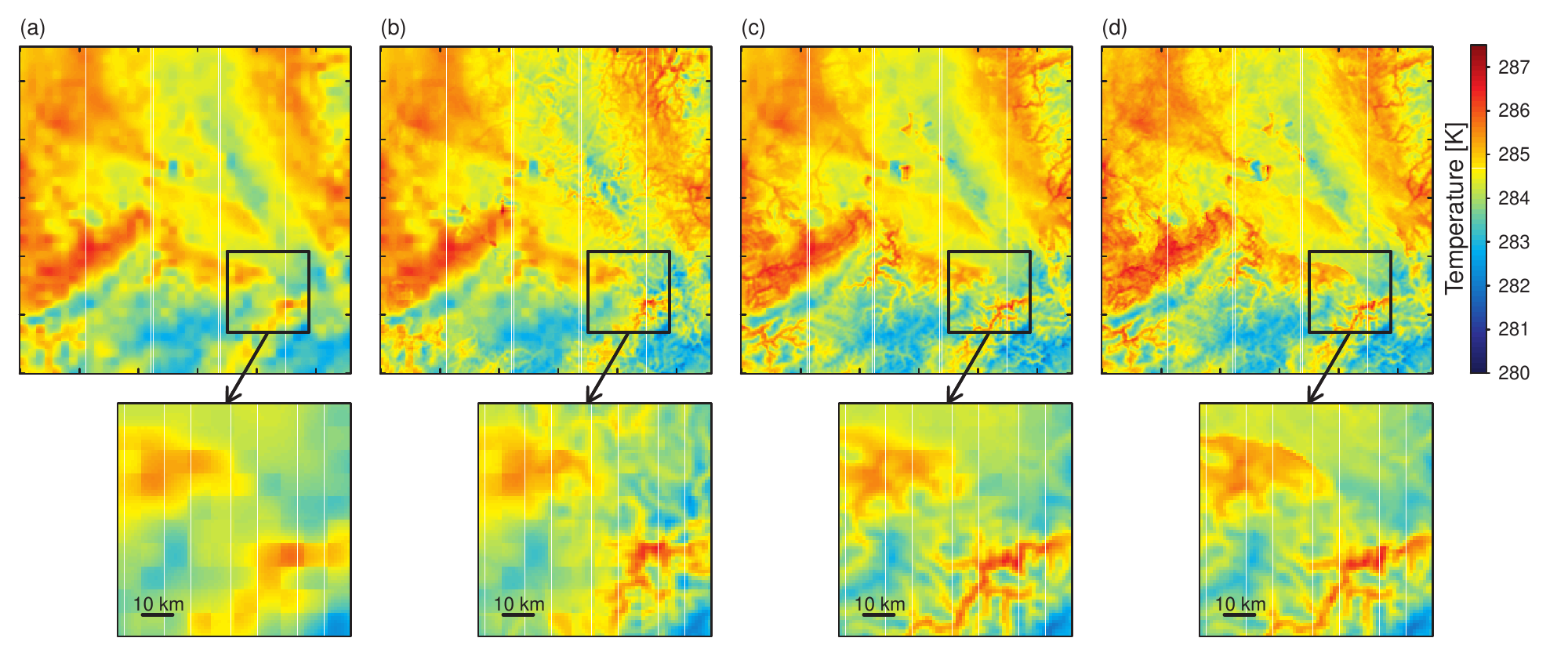}
	\caption[]{Performance of a good solution (rating visually) validated on October 14th 2007.
		   Shown is the temperature field on the validation day at 11:00 UTC.
		   The upper figures show the full fields (i.e., 112 $\times$ 112 km); the bottom figures show a zoom in on an area of 28 $\times$ 28 km:
		   (a) shows the spline-interpolated field; (b) shows the field resulting from the downscaling rule by \cite{schomburg2010downscaling}; (c) shows the field resulting from the rule found by GP; (d) shows the reference field from the high-resolution model run.}
\label{fig:fieldday}
\end{figure*}

\begin{figure*}
\centering
	
	\includegraphics[width=0.9\textwidth]{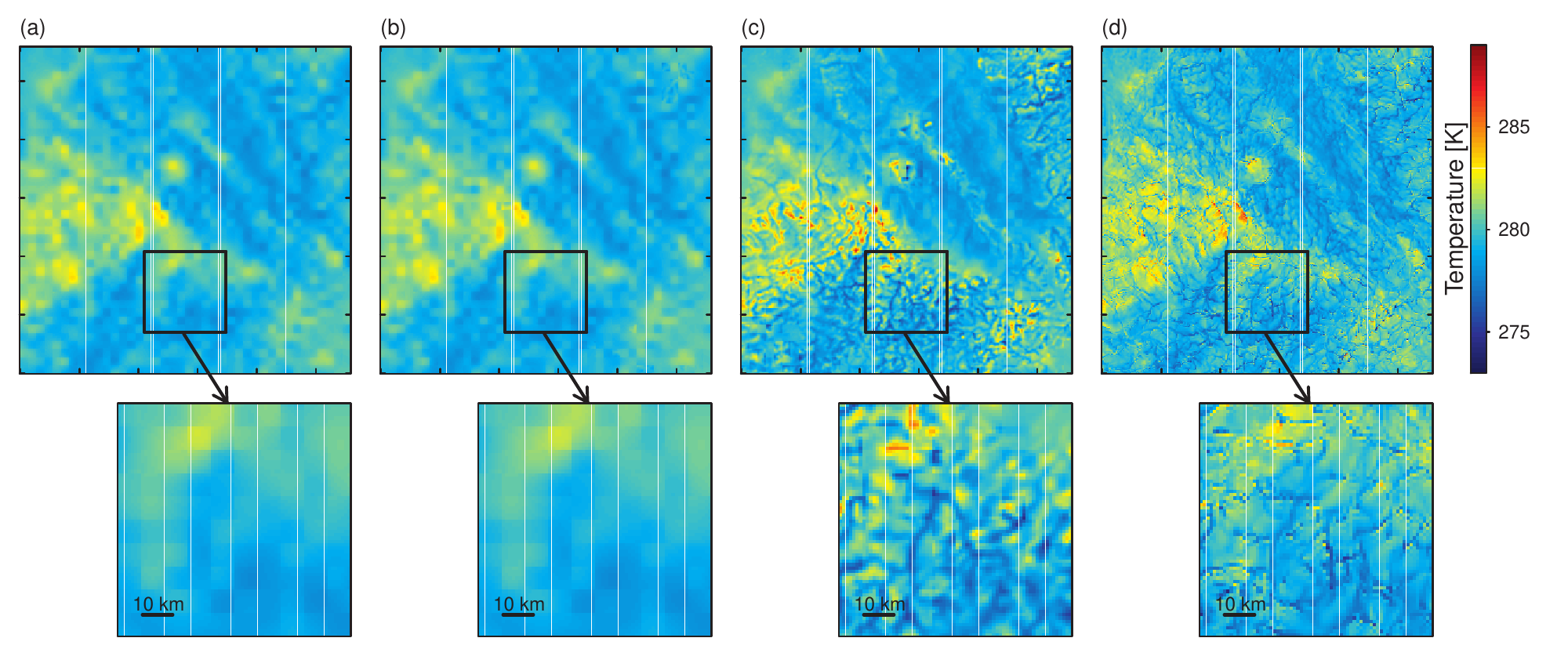}
	\caption[]{Same as in Figure \ref{fig:fieldday}, but for October 14th 2007 23:00 UTC.}
\label{fig:fieldnight}
\end{figure*}

\begin{figure*}
\centering
	\includegraphics[width=0.5\textwidth]{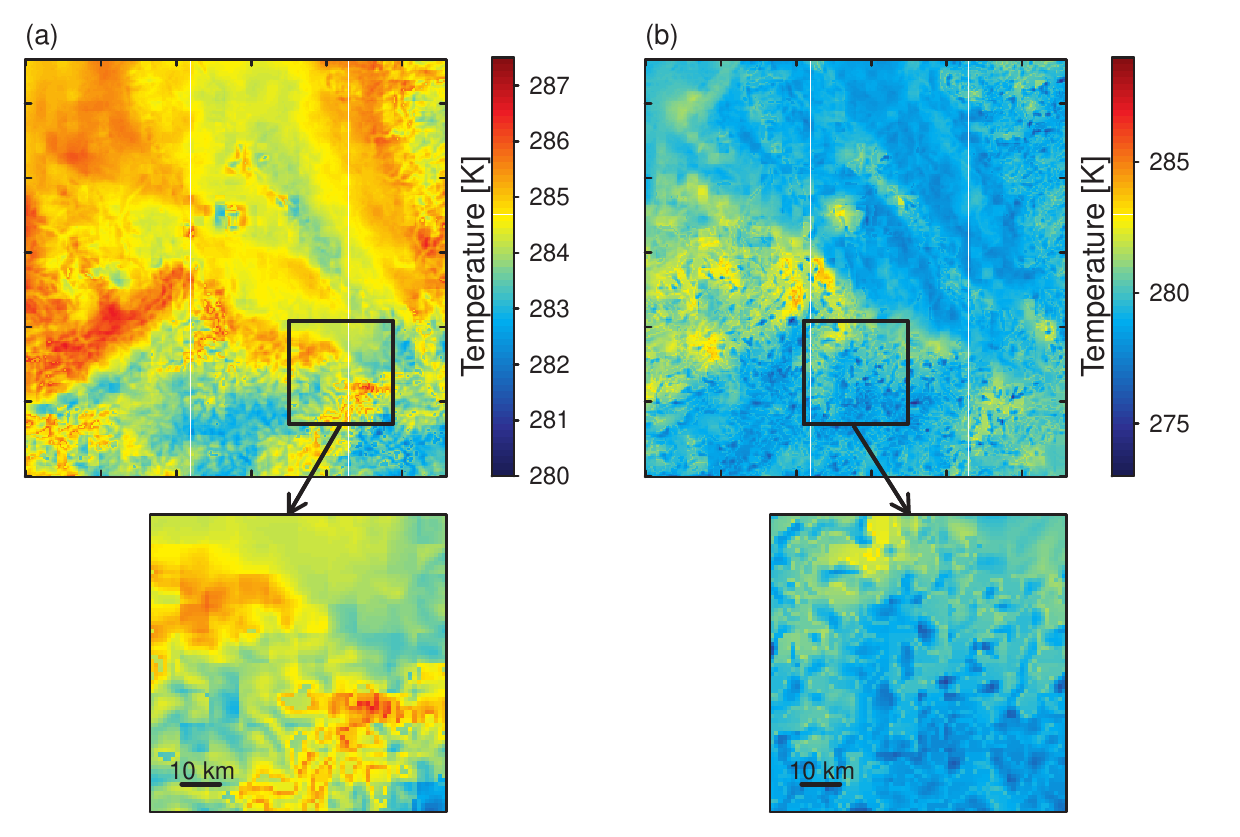}
	\caption[]{Performance of an average solution (rating visually) validated on October 14th 2007:
		   (a) shows the downscaled field at 11:00 UTC (see Figure \ref{fig:fieldday});
		   (b) shows the downscaled field at 23:00 UTC (see Figure \ref{fig:fieldnight}).}
\label{fig:fieldday2}
\end{figure*}